\documentclass[superscriptaddress,showpacs,amsmath,amssymb,aps,showkeys,floatfix,prd,onecolumn,a4paper,preprint]{revtex4-1}
\usepackage{graphicx}
\usepackage{epstopdf}
\usepackage{dcolumn}
\usepackage{bm}
\usepackage{epsfig}
\usepackage{amsfonts}
\usepackage{amssymb,amscd}
\usepackage{hyperref}
\usepackage{xcolor}
\hypersetup{
    colorlinks=true,                
    breaklinks=true,                
    urlcolor= black,                
    linkcolor= blue,                
    bookmarksopen=false,
    filecolor=black,
    citecolor=blue,
    linkbordercolor=blue,
    pdfborder = {0 1 0}
}

\begin{document}

\title{Prompt photon production in high-energy $pA$ collisions at forward rapidity}

\pacs{12.38.-t; 13.60.Le; 13.60.Hb}

\author{G. Sampaio dos Santos}
\affiliation{High Energy Physics Phenomenology Group, GFPAE  IF-UFRGS \\
Caixa Postal 15051, CEP 91501-970, Porto Alegre, RS, Brazil}

\author{G. Gil da Silveira}
\affiliation{High Energy Physics Phenomenology Group, GFPAE  IF-UFRGS \\
Caixa Postal 15051, CEP 91501-970, Porto Alegre, RS, Brazil}
\affiliation{Departamento de F\'{\i}sica Nuclear e de Altas Energias, Universidade do Estado do Rio de Janeiro\\
CEP 20550-013, Rio de Janeiro, RJ, Brazil}

\author{M. V. T. Machado}
\affiliation{High Energy Physics Phenomenology Group, GFPAE  IF-UFRGS \\
Caixa Postal 15051, CEP 91501-970, Porto Alegre, RS, Brazil}

\begin{abstract}

Prompt photon production in hadronic collisions at the RHIC and the LHC energies is investigated within the QCD color dipole approach. Predictions for the nuclear modification factor in $pA$ collisions are evaluated based on parton saturation framework and the results are compared to the experimental measurements as a function of the photon transverse momentum at different rapidity bins. The reliability of the models is performed with the data from PHENIX, ATLAS, and ALICE Collaborations. Moreover, we show that the observed $x_T$-scaling of prompt photon production in $pp$ and $pA$ collisions can positively be addressed in the QCD color dipole formalism. 
\end{abstract}

\maketitle

\section{Introduction}
\label{intro}

In high-energy collisions involving a nuclei, the presence of effects associated to the nuclear environment modify the behavior of the partonic distributions. A detailed understanding of the initial- and final-state effects, associated to the stages of the collision, is crucial to describe the data from heavy-ion collisions (HIC) at the RHIC and the LHC. The suitable scenario to analyze such effects would be the nuclear deep inelastic scattering (nDIS), which is the plan of the future Electron-Ion Collider \cite{EIC}. Alternatively, proton-nucleus ($pA$) collisions can be used as a probe of the nuclear effects, since the formation of a quark-gluon deconfined medium known as Quark-Gluon Plasma (QGP) is not expected in this case. For a better understanding of the scenario created in $AA$ reactions, $pA$ collisions can be used as a baseline to disentangle the initial- and final-state effects. Hence, one needs to evaluate such effects before testing the signals from high density QCD medium that can be identified in $AA$ collisions. Consequently, a consistent knowledge of the measurements in $pA$ collisions is essential to improve the comprehension of the underlying physics in HIC. Usually, analyzing the nuclear effects is made by measuring a nuclear modification factor, which can establish a reference for the collision centrality or system-size dependence. At the RHIC  energy \cite{arsene, adams, adare}, it was observed a suppression for pion production in $dAu$ collisions and such a particular result is an important source to constraint the nuclear parton distribution function (nPDF). At the LHC, investigations about nuclear modification factor for $\pi^0$ and the ratio of prompt photons to pion production, $\gamma/\pi^0$, have been used to verify the self-consistency of the QCD approaches (see, for instance, discussions in Refs.~\cite{Goncalves:2020tvh,JalilianMarian:2012bd,amir}).

Here we focus on an important hard probe of nuclear environment, namely the production of hard isolated photons. At the high-energy regime, the nucleus target is probed at small Bjorken variable $x$, and such kinematic region can be accessed on measurements of prompt photons at forward rapidities. Measurements of prompt photon cross sections have been proposed as a clean source of information about the QCD dynamics \cite{pasechnik,acharya, gordon, frixione}. Due the nature of the quark-photon vertex, the only interaction is electromagnetic, especially because photons are colorless probes of the dynamics of quarks and gluons. Also, direct photons are not disturbed by final interactions, then they can leave the system without loss of energy and momentum. Other useful property is the elementary diagrams for the underlying processes, which are theoretically well established and the contribution from fragmentation processes can be suppressed by an isolation criteria. Studies of nPDFs using prompt photons have been proposed in Ref.~\cite{helenius1}, demonstrating that experimental data on this process can strongly constrain them. In particular, gluon distribution, which are not well constrained at small-$x$ and there are large theoretical uncertainties from usual perturbative QCD (pQCD), can be extracted in a precise way. Towards to low values of $x$ the gluon density substantially increases, bringing concerns about unitarity violation. At the low-$x$ regime the growth of the gluon density can be controlled by gluon recombination effect, which is a nonlinear QCD phenomenon leading to the gluon saturation, it is expected that the low-$p_T$ photon distribution can probe this dense and saturated regime. 

The treatment of the prompt photon production can be developed within the QCD color dipole (CD) formalism, where the production mechanism resembles a bremsstrahlung \cite{kop1,kop2}. The photon emission is viewed as a quark/antiquark electromagnetic bremsstrahlung, which exchange a single gluon with the target \cite{kop}. Hence, one can interpret the real photon radiation process in terms of $q\bar{q}$ dipole scattering off the target. The main ingredient in the CD approach is the universal dipole cross section, fitted to DIS data and successfully describes the DESY-HERA $ep$ data for inclusive and exclusive processes. The dipole cross section takes into account the nonlinear gluon recombination effect that is expected to be relevant at low $x$. In the parton saturation picture, a scaling property associated to the DIS takes place, namely geometric scaling phenomenon. The cross sections for photon-target processes are function of a dimensionless single scaling variable \cite{munier}, instead of two independent variables, such as $x$ and $Q^2$ (photon virtuality). Such a property can be extended to single particle production in hadron-hadron or $p(d)A$ collisions. We will show it can explain the $x_T$ scaling observed in prompt photon production in $pp$, $dA$, and $pA$ reactions at central rapidities.

In this work, predictions are done for the nuclear modification factor considering the RHIC and LHC kinematic regimes. Direct photon production at large- and low-$p_T$ in a wide rapidity range is considered. These results are an extension of the previous investigations presented in Ref.~\cite{gsds}, where the differential cross section in $pp$ and $pA$ collisions at the LHC energies has been analyzed. Moreover, we carefully examine the theoretical mechanism responsible for the observed $x_T$-scaling in $pp/pA$ collisions. 

The paper is organized as follows. In Sec.~\ref{dirph} the theoretical framework is presented, including the main expressions used in our calculations within the CD formalism. In Sec.~\ref{res} we show our theoretical results, discussing and comparing them to the measurements available at the RHIC and the LHC. The last section presents the main conclusions and remarks.

\section{Theoretical formalism}
\label{dirph}

The nuclear modification factor $R_{pA}$ is determined as the ratio of $pA$ to $pp$ cross sections properly scaled with the correspondent mass number $A$ of the target nucleus,
\begin{eqnarray}
R^{\gamma}_{pA}(y,p_T) = \frac{d^3\sigma(pA \rightarrow \gamma A)/dy^{\gamma}d^2\vec{p_T}}{A \cdot d^3\sigma(pp \rightarrow \gamma p)/dy^{\gamma}d^2\vec{p_T}}.
\label{ratio} 
\end{eqnarray}
The advantage in using $R_{pA}$ consists in the cancellation of uncertainties that came from the individual cross sections in the ratio.
The differential cross section for prompt photon production in $pp$ collisions in terms of the photon rapidity $y^{\gamma}$ and transverse momentum $p_{T}$ was derived in Ref.~\cite{kop3} and is written as
\begin{eqnarray}
\frac{d^3\sigma\,(pp\to \gamma
X)}{dy^{\gamma}d^{2}\vec{p}_{T}} & = &
\frac{\alpha_{em}}{2\pi^2}\int_{x_{1}}^{1}\frac{d\alpha}{\alpha}
 F_{2}^{(P)}\left(\frac{x_{1}}{\alpha},\mu^2\right)  \left\{ m_q^2\alpha^4\left[\frac{{\cal I}_1}{(p_T^2+\varepsilon^2)}-\frac{ {\cal I}_2}{4\varepsilon} \right]
  +  [1+(1-\alpha)^2]\right.\nonumber \\
  &\times& \left. \left[ \frac{\varepsilon p_T \, {\cal I}_3}{(p_T^2+\varepsilon^2)} -\frac{{\cal I}_1}{2}+\frac{\varepsilon \,{\cal I}_2}{4}\right]
\right \},
\label{hank}
\end{eqnarray}
where $F_{2}^{(P)}$ stands for the structure function for the projectile ($P$) particle and ${\cal I}_{1,2,3}$ are Hankel integral transforms of order $0$ (${\cal I}_{1,2}$) and order $1$ (${\cal I}_{3}$) given by
\begin{eqnarray}
\label{hankel1}
{\cal I}_1 & = & \int_0^{\infty}dr\,rJ_0(p_T\,r)K_0(\varepsilon\,r)\,\sigma_{dip}(x_2,\alpha r),\\
{\cal I}_2  &=&  \int_0^{\infty}dr\,r^2J_0(p_T\,r)K_1(\varepsilon\,r)\, \sigma_{dip}(x_2,\alpha r), \\
{\cal I}_3 & = & \int_0^{\infty}dr\,rJ_1(p_T\,r)K_1(\varepsilon\,r)\, \sigma_{dip}(x_2,\alpha r).
\label{hankel3}
\end{eqnarray}
In numerical calculations we will consider a $F_{2}^{(P)}$ parametrization given in Ref.~\cite{adeva} (for proton and deuterium) and $\mu^{2} = p^{2}_{T}$. The choice of the scale $\mu^2$ is one of the theoretical uncertainties in the formalism. Moreover, the fraction of the quark momentum carried by the photon is denoted by $\alpha$ and momentum fractions $x_{1,2}$ have the form $x_{1,2} = \frac{p_T}{\sqrt{s}}e^{\pm y^{\gamma}}$, where $\sqrt{s}$ is the collision center-of-mass energy. In the Hankel transforms, an effective quark mass appears in the auxiliary variable $\epsilon^{2}=\alpha^{2} m_{q}^{2}$, which is taken as $m_{q}=0.2$~GeV in our calculations.

Another quantity that enters in the Hankel transforms is the dipole cross section $\sigma_{dip}$, a crucial ingredient to perform a calculation that can be compared to experimental measurements. Common features presented by $\sigma_{dip}$ are: (i) it saturates for large dipole transverse sizes, $r$, i.e., $\sigma_{dip} \rightarrow \sigma_0$; (ii) for small dipole sizes the dipole cross section behaves like $\sigma_{dip} \sim r^2$, i.e., vanishes accordingly with the color transparency phenomenon \cite{kop4}. Here, models for dipole cross sections based on the idea of gluon saturation and constrained by recent data available from $ep$ collisions at DESY-HERA collider will be used. Explicitly, the following parametrizations will be considered: the GBW model \cite{gbw}, with more recent fitting parameters reported in Ref.~\cite{gbwfit}, and the IPSAT model \cite{ipsat}, where the parameters are given in Ref.~\cite{ipsatfit} and such approach includes QCD gluon evolution via DGLAP equation. It should be noticed that in the color transparency regime the Hankel integrals can be analytically performed \cite{mm}. We will discuss this case in detail when the $x_T$ scaling is studied.

For a heavy target, nuclear effects are related to multiple parton scattering as well as nonlinear gluon recombination. We employ the state-of-art of phenomenological models to the dipole-nucleus amplitudes $N_A$, which contain explicit impact parameter dependence or geometric scaling. There are basically two ways to implement the nuclear effects within the CD approach: (i) geometric scaling (GS) property from parton saturation models; and (ii) Glauber-Gribov (GG) formalism for nuclear shadowing. First, we follow Ref.~\cite{salgado} to apply the GS including the $A$-dependence in the scattering cross section. There, the authors have demonstrated that the nuclear DIS cross section at small-$x$ is directly associated to the cross section for DIS off proton target. Hence, the proposed GS assumes that the nuclear effects are absorbed into the saturation scale and on the nucleus transverse area, $S_A=\pi R_A^2$, compared to the proton case, $S_p=\pi R_p^2$. Consequently, the saturation scale in protons, $Q_{s,p}$, is replaced by a nuclear saturation scale, $Q_{s,A}$, which is translated into an $A$-dependence, 
\begin{eqnarray}
Q_{s,A}^2&=&Q_{s,p}^2\left(\frac{A \pi R_p^2}{\pi R_A^2}\right)^{\Delta}, \label{qs2A} \\
N_A(x,r,b) & = & N(rQ_{s,p}\rightarrow rQ_{s,A}), \label{NA}
\end{eqnarray}
which grows with the quotient $\Delta = 1+\xi$ with $\xi=[(1-\delta)/\delta]$. Here, $R_A \simeq 1.12 A^{1/3}$~fm is the nucleus radius, whereas the quantities $\delta = 0.79$ and $\pi R_p^2=1.55$~fm$^2$ have been determined by data \cite{salgado}. The prompt photon production cross section in $pA$ is rescaled accordingly as follows,
\begin{eqnarray}
\frac{d^3\sigma(pA\to \gamma X)}{dyd^2\vec{p}_T} = \left(\frac{S_A}{S_p}\right)\left.\frac{d^3\sigma(pp\to \gamma X)}{dyd^2\vec{p}_T}\right|_{ Q_{s,p}^2\rightarrow Q_{s,A}^2}.
\label{prescr}
\end{eqnarray}
We mention Ref.~\cite{ben} where the GS property has been employed in order to describe data for both the DVCS at DESY-HERA and the exclusive meson production in DESY-HERA and LHC colliders. 

Otherwise, in terms of GG formalism which includes the multiple elastic scattering diagrams related to the dipole-nucleus interaction, the nuclear scattering cross section is written as \cite{armesto},
\begin{eqnarray} 
\sigma_{dip}^{nuc} (x,\vec{r};A) &=& 2\int d^2b \,\left\{1-\exp\left(-\frac{1}{2}\sigma_{dip}(x,r)T_A(b)\right)\right\},
\label{ipsat2}
\end{eqnarray}
with $\sigma_{dip}$ being the dipole-proton cross section and $T_A$ is the nuclear profile function obtained from the Woods-Saxon distribution. Such a model was considered in Ref.~\cite{armesto}, showing results in good agreement with the existing experimental data on the ratios of nuclear structure functions, $F_2^A/F_2^B$. 

Still on the color transparency regime within the CD picture, a scaling property for the invariant cross section of prompt photon in $pp/pA$ collisions on the variable $x_T = 2\,p_T/\sqrt{s}$ (the so-called $x_T$-scaling) can be derived. Taking the massless limit, $m_q \rightarrow 0$ in Eq.~(\ref{hank}), the second term holds, with the only contribution that survives from the Hankel integrals being proportional to an analytic function, ${\cal I}_1\propto\sigma_0 (\alpha Q_s)^2/p_T^4$. This last result is a consequence of considering the color transparency in the dipole-target cross section. Furthermore, a rough approximation can be obtained for the nucleon structure function assuming the GBW model (with $\gamma_s =1$),
\begin{eqnarray}
F_2(x,Q^2)\approx\frac{\sigma_0 Q^2}{4\pi^2\alpha_{em}}\left(\frac{Q_s^2(x)}{Q^2}\right)^{\gamma_s},
\end{eqnarray}
where the saturation scale is set as $Q_s^2(x)=Q_0^2(x_0/x)^{\lambda}$ (with $Q_0=1$~GeV and parameters $x_0$ and $\lambda$ being fitted from HERA data at small-$x$).
Hence, taking into account the assumptions established above and further integrating Eq.~(\ref{hank}) over $\alpha$, a $x_T$-scaling expression is obtained for the $pp$ case,
\begin{eqnarray}
E \frac{d^{3}\sigma^{pp\to \gamma X}}{d^3p}(x_T) & \approx & \frac{N_0}{\big(\sqrt{s}\big)^{4}}\,\bigg(\frac{x_T}{2}\bigg)^{-n}\,f(x_1),
\label{xtpp}
\end{eqnarray}
with $n=2\lambda+4\simeq 4.5$ and $f(x_1)\approx (1012/1989) -(4/17)x_1^{17/4}+(8/13)x_1^{13/4}-(8/9)x_1^{9/4}$ being a well behaved function of $x_1=(x_T/2)e^y$ resulting from the $\alpha$-integration. Moreover, the overall normalization is given by $N_0 = \bar{\sigma}_{pp}\,(x_0)^{2\,\lambda}$, with parameters 
$\bar{\sigma}_{pp} = 0.035$~mb/GeV$^2$, $x_0 = 0.4\times 10^{-4}$ and $\lambda = 0.248$ taken from GBW model. On the other hand, for $pA$ collisions based on GS proposed in Ref.~\cite{salgado}, the invariant cross section reads 
\begin{eqnarray}
E \frac{d^{3}\sigma}{d^3p}(pA\to \gamma X) & \approx & \frac{N_0}{\big(\sqrt{s}\big)^{4}}\,\bigg(\frac{S_A}{S_p}\bigg)\,
\bigg(\frac{A\,S_p}{S_A}\bigg)^{\Delta}\,\bigg(\frac{x_T}{2}\bigg)^{-n}\,f(x_1), \\
& \approx & A\bigg(\frac{A\,S_p}{S_A}\bigg)^{\xi}E \frac{d^{3}\sigma}{d^3p}^{pp\to \gamma X}(x_T),
\label{xtpA}
\end{eqnarray}
where $\xi=(1-\delta)/\delta \simeq 0.27$. The value of $\bar{\sigma}_{pp}$ is determined in order to describe the  lower energy data in $pp$ collisions and we set the same value for $pA$ reactions. The simple parametrization presented above can be further sophisticated by leaving the anomalous dimension, $\gamma_s$, as a free parameter or using a $p_T$-dependence like in the BUW model \cite{buw}.  Similar proposals of scaling can be found in Refs.~\cite{kbml, pras, kp}, where the scaling observed in prompt photon production is related to an universal multiplicity scaling. The latter is studied using the charged hadron pseudorapidity density at midrapidity, $dN_h/d\eta$. 

In the next section a comparison is performed between the theoretical approach based on QCD dipole picture and experimental data from RHIC and LHC colliders.

\section{Results and discussions}
\label{res}

In this section, we present the numerical calculations concerned the nuclear modification factor $R_{pA}$ obtained with the CD approach, where we used the GBW and IPSAT phenomenological models
for the dipole cross section. We investigate the influence of nuclear effects in low and large-$p_T$ prompt photon production via GS and GG formalism. Some comments are in order here. We compute the dipole-nucleus amplitude considering the GBW model as an input for GS and GG implementations. In our calculations with the IPSAT model we are applying the small-$r$ limit for $\sigma_{dip}$, which is appropriate at large $p_T$ domain given that $r \approx 1/p_T$ in direct photon production and also enables us to analytically solve the Hankel transforms discussed in the previous section.
Furthermore, there is no significant change regarding the option for the proton structure function in Eq.~(\ref{hank}). It has been verified that employing $F^{p}_2$ from Ref.~\cite{adeva} or the ALLM2007 parametrization \cite{allm}, the numerical results are nearly identical. As a last remark, the CD approach has a threshold of validity taken as $x_2\leq10^{-2}$, which is, in principle, well suitable for small $x_2$. However, in Ref.~\cite{mm} is demonstrated that a large-$x$ correction should be added to consistently describe the prompt photon phenomenology. Therefore, we have multiplied the GBW dipole cross section by a threshold factor $(1-x_2)^n$ (with $n=7$). In the IPSAT model, the parametrization for the gluon PDF already contains the threshold factor. 

\begin{figure*}[t]
\begin{tabular}{cc}
\includegraphics[scale=0.35]{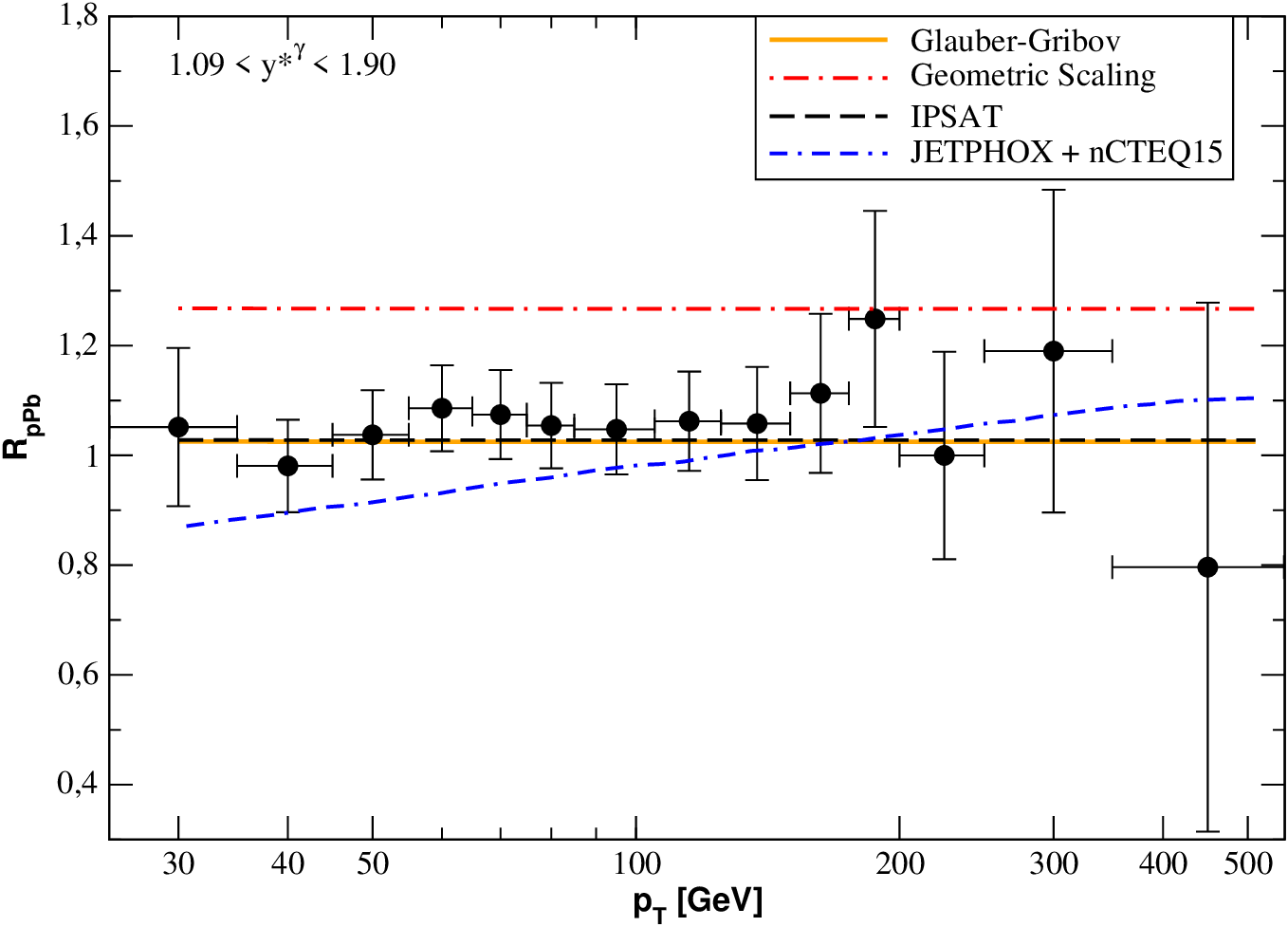}
\includegraphics[scale=0.35]{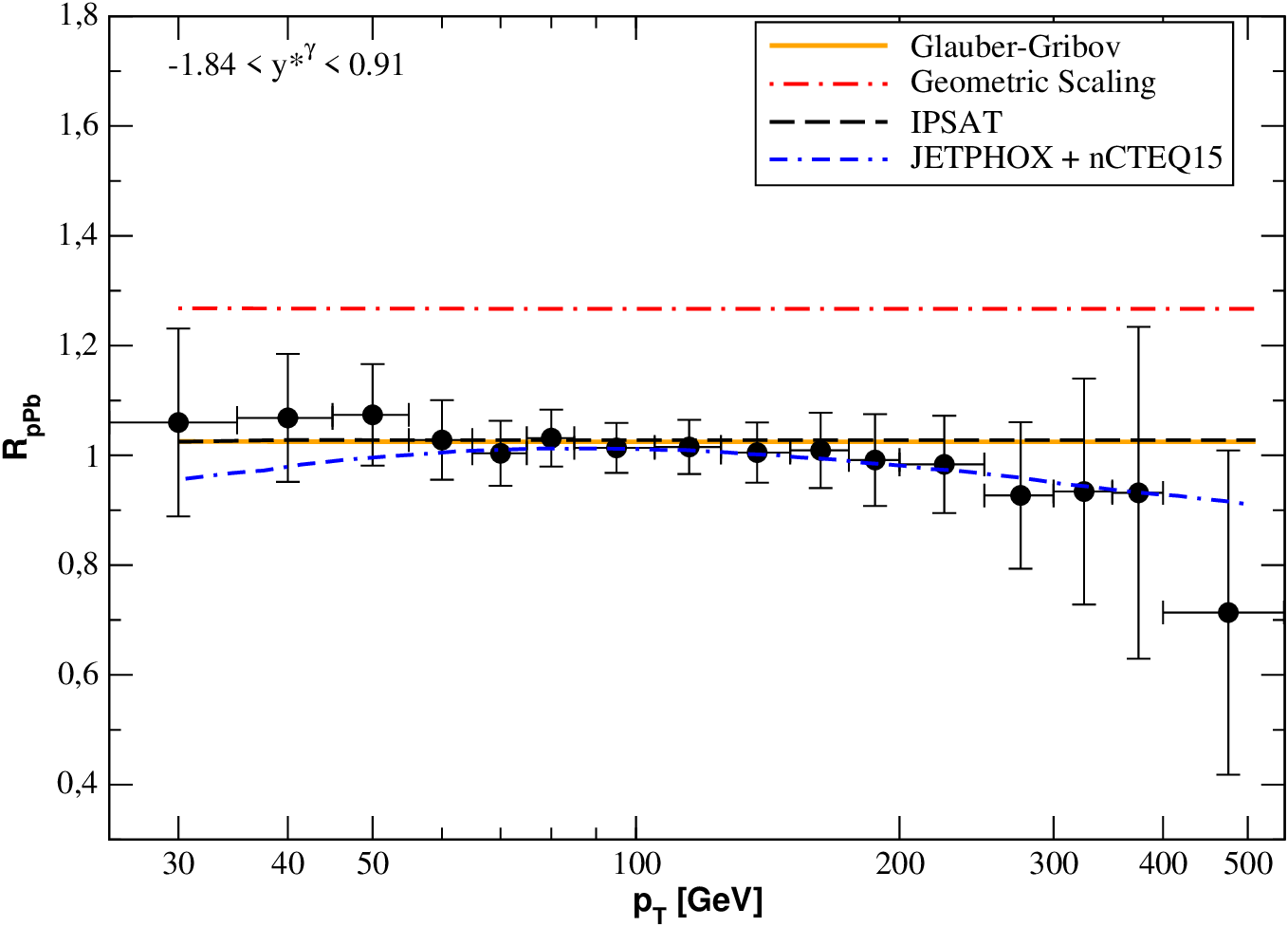}
\end{tabular}
\caption{Nuclear modification factor $R_{pPb}$ as a function of $p_T$ shown for two forward rapidity bins at 
$\sqrt{s} = 8.16$~TeV. The predictions are obtained using GG, GS, and IPSAT models. Results from JETPHOX Monte Carlo using the nPDF nCTEQ15 are also presented as a matter of comparison, together with experimental data from the ATLAS Collaboration \cite{aaboud1}.}
\label{rpA816}
\end{figure*}

Now, in Fig.~\ref{rpA816} we show the results for the nuclear modification factor in $pPb$ collisions at $\sqrt{s} = 8.16$~TeV compared to the measurements by the ATLAS experiment \cite{aaboud1} as a function of $p_T$ and $y^{*\gamma}$.
Moreover, the results with pQCD at NLO level of direct and fragmentation contributions to the cross-sections using the JETPHOX Monte Carlo \cite{jetphox} and nCTEQ15 nuclear PDF \cite{kovarik} are also included in order to perform a comparison with our results. 
Considering the two rapidity bins, the measured nuclear modification factor is consistent with unity, indicating that the magnitude of nuclear effects becomes negligible. In addition, the GG and IPSAT approaches predict quite small nuclear effect, while the GS model predict $R_{pA} \gtrsim 1$. Concerning the GS approach, the nuclear ratio has a form [see Eqs.~(\ref{xtpp}) and (\ref{xtpA})],
\begin{eqnarray}
R^{\gamma}_{pA}\approx\left( \frac{A\pi R_p^2}{\pi R_A^2} \right)^{\xi},
\end{eqnarray}
with $\xi \simeq 0.27$. Accordingly, this reproduces numerically $R_{pPb}\simeq 1.3$ for any value of $p_T$. For the IPSAT case, the small $r$ approximation allows us to expand the eikonalized amplitudes as $N_p\approx (\pi^2\alpha_s/2N_c)r^2 xgT_p(b)$ and $N_A\approx (\pi^2\alpha_s/2N_c)r^2 xgT_A(b)$. Besides, the normalization of the proton and nuclear thickness function, $\int d^2\vec{b}\,T_A(b)=A$ ($A=1$ for proton case), implies that $\sigma_{dip}^{nuc}=A\sigma_{dip}$ and results in $R_{pPb}\approx 1$. The GG and IPSAT results are fairly similar to those from JETPHOX Monte Carlo with nCTEQ15. We will see that the situation changes in the low $p_T$ case.

\begin{figure*}[t]
\begin{tabular}{cc}
\includegraphics[scale=0.35]{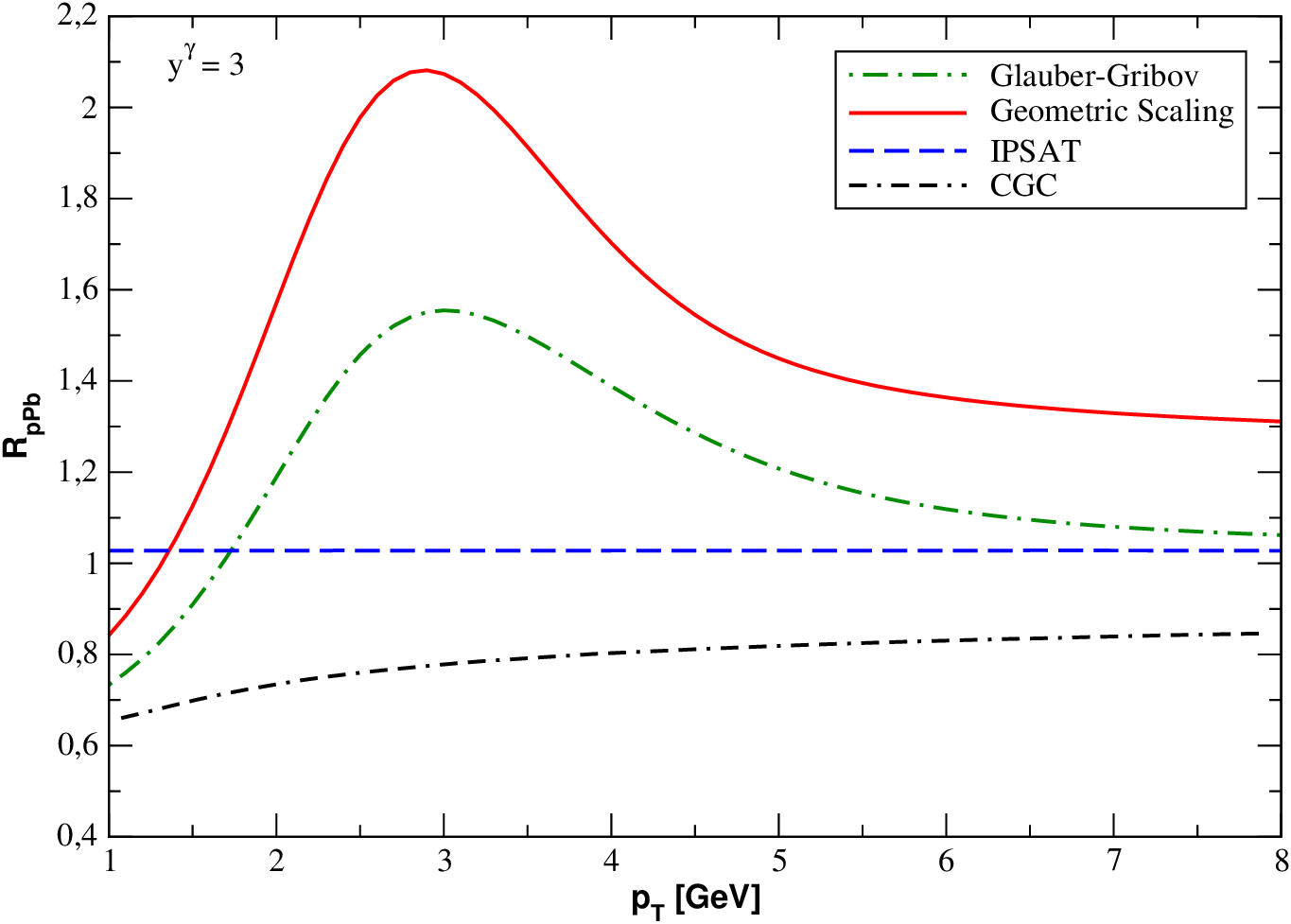}
\includegraphics[scale=0.35]{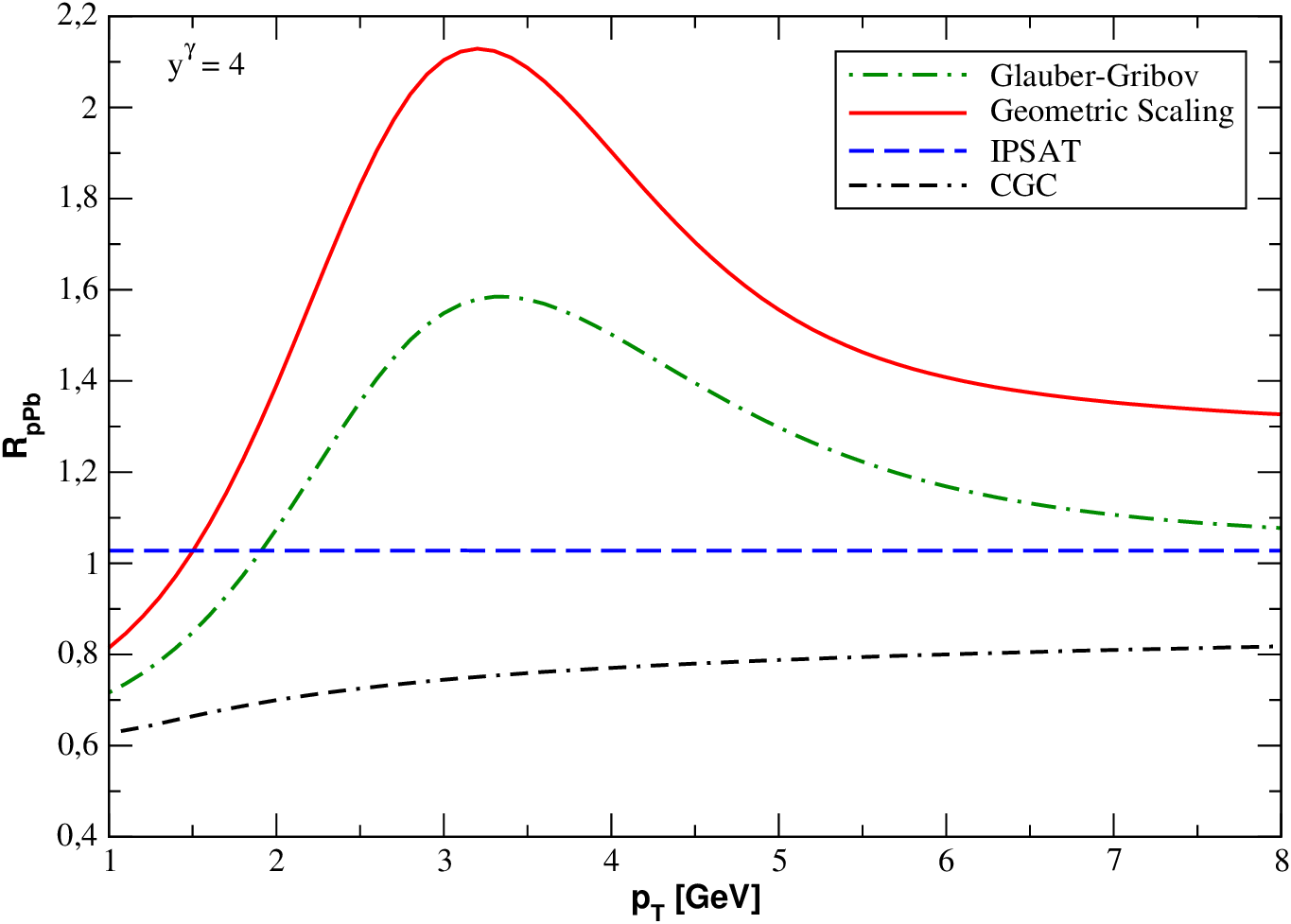} \\
\includegraphics[scale=0.35]{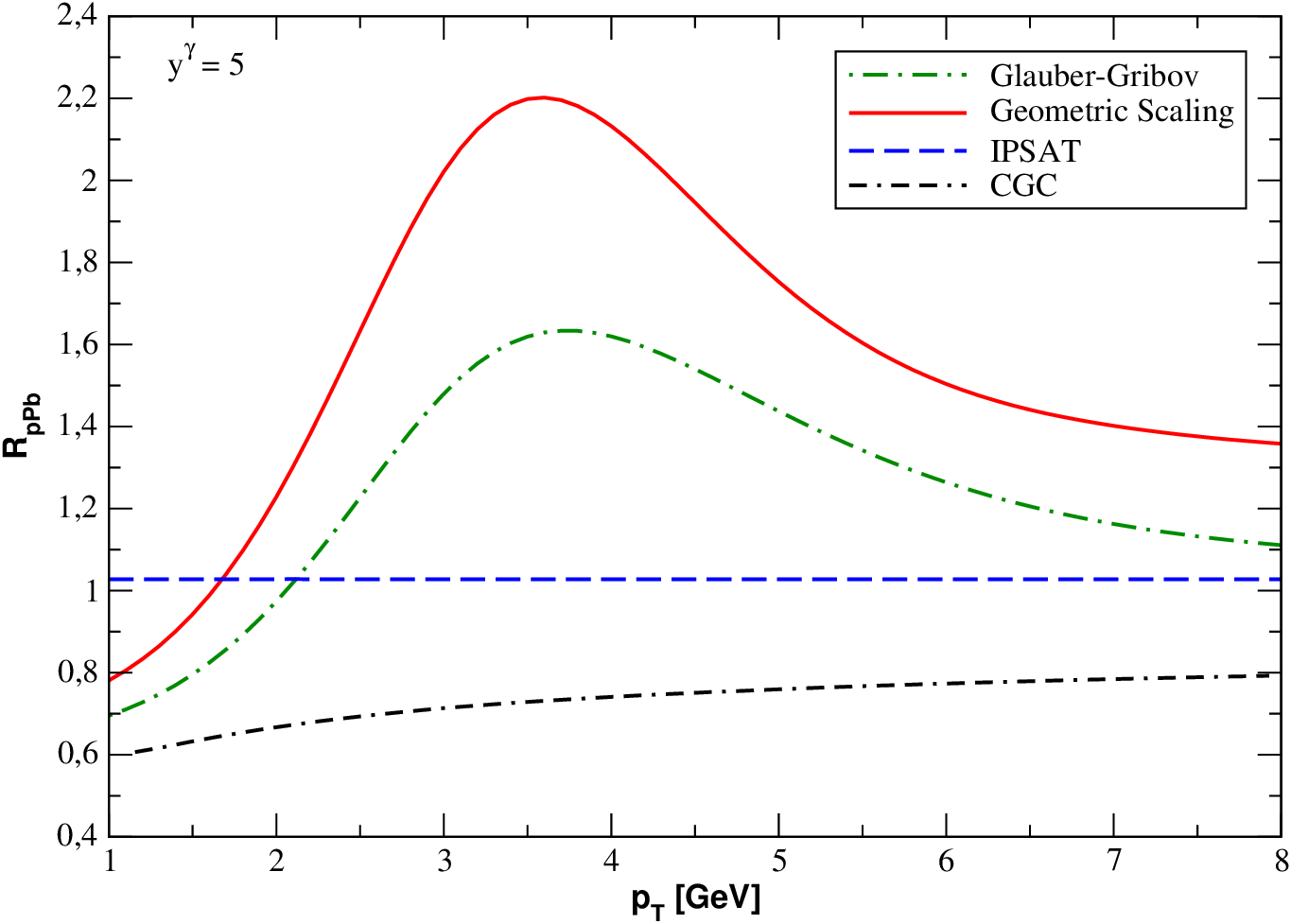}  
\end{tabular}
\caption{Nuclear modification factor $R_{pPb}$ for prompt photon as a function of $p_T$ shown for different fixed values of the photon forward rapidity at $\sqrt{s} = 8$~TeV. The predictions are obtained using GG, GS, and IPSAT approaches and compared to the results from CGC effective field theory.}
\label{compPb}
\end{figure*}

In order to test the nuclear effects that have been addressed, our predictions for $R_{pA}$ are compared to recent studies in the literature employing others approaches. We start discussing Ref.~\cite{ducloue}, where calculations are performed considering $pPb$ collisions at energy of 8~TeV and based on CGC formalism using CD cross sections solved from the running coupling Balitsky-Kovchegov (BK) evolution equation. The CGC formalism predicts a consistent suppression at forward rapidities in the range $1\leq p_T\leq 8$~GeV. In Fig.~\ref{compPb} we present the $R_{pA}^{\gamma}$ predictions at low-$p_T$ for fixed values of the photon forward rapidity, $y^{\gamma} =$ 3, 4, and 5, respectively.
Up to $p_T \approx 2$~GeV, GG, GS, and CGC models predict similar results with a suppression pattern. This is well understood in terms of the nuclear saturation scale. At low transverse momentum and forward rapidities, small $x_2=(x_T/2)e^{-y}$ is probed. For instance, at $p_T=2$~GeV and $y^{\gamma}=4$, one has $x_2\sim 6.6\times 10^{-6}$ and the proton saturation scale reaches $Q_{s,p}^2\simeq 1.7$~GeV$^2$ and the corresponding nuclear saturation scale $Q_{s,Pb}^2\approx 3Q_{s,p}^2\simeq 5$~GeV$^2$. As we can see, one has $p_T^2\leq Q_{s,A}^2$ at low $p_T$ and at forward rapidities at the LHC, expecting an important shadowing correction. As $p_T$ increases at fixed rapidity, a transition from saturated to dilute regime is reached, $p_T^2\gg Q_{s,A}^2$, and nuclear corrections are weaker. An enhancement of $R_{pPb}$ is verified as excepted for the IPSAT model. However, the results with GG and CGC tend to be closer to unity at $p_ T \geq 8$~GeV in contrast with GS that continues showing an enhancement of $R_{pPb}$. A Cronin enhancement has been observed in both GG and GS results. Moreover, the location of the peak depends on the rapidity and it moves into the direction of larger $p_T$ in accordance with the increase of the rapidity. The peaks have the same shape in both predictions and differ in their height.

The Cronin-like peak is typical in models including rescattering, mostly at midrapidities. In Ref.~\cite{amir}, this issues has been first investigated and it was found that the Cronin enhancement can survive at the LHC energy within the saturation QCD dipole models. We observe the same pattern for GG and GS calculations. This is consistent with recent studies on pion-photon correlations presented in Ref.~\cite{Goncalves:2020tvh}. In Ref.~\cite{amir}, the peak height can be reduced if gluon shadowing, $R_G$ (from $|q\bar{q}g\rangle$ Fock state contribution), in the form $\sigma_{dip}\rightarrow R_G(x_2)\sigma_{dip}$ enters in Eq.~(\ref{ipsat2}). Namely, a subtle cancellation between the saturation and gluon shadowing effects can leads to a rather small Cronin peak in $R_{pA}^{\gamma}$. It is clear that further experimental analyzes of the ratio $R_{pA}^{\gamma}$ at very forward $y^{\gamma}$ would be very fruitful to draw a distinction between the CGC and its competing approaches.

The discussion above can be placed in juxtaposition with results from usual pQCD approach. At leading order (LO) accuracy, it can be shown \cite{Arleo:2007js} for central rapidities, $y^{\gamma}\approx 0$, that $R_{pA}^{\gamma}(x_T)\simeq \frac{1}{2}[R_{F_2}^A(x_T)+R_G^A(x_T)]$. That is to say the nuclear modification factor is a linear combination of nuclear ratio for gluons, $R_G$, and nuclear ratio for structure functions in a nucleus $A$ and at midrapidity both contributions have similar weights. On the other hand, at forward rapidities, $y^{\gamma}>y_c^{\gamma}$ (let us say $y_c^{\gamma}\sim 2$), the relation becomes $R_{pA}^{\gamma}(x_T, y^{\gamma})\simeq R_G^A(x_Te^{-y^{\gamma}})$ and the pQCD approach is quite close to ours.
The relations between nuclear factors and the nuclear gluon PDF/nuclear structure functions were shown to remain as a correct approximation up to a few-percent accuracy in calculations in next-to-leading (NLO) order level \cite{Arleo:2007js}. We quote Refs.~\cite{helenius1,Goharipour:2017uic,Goharipour:2018sip,Klasen:2017dsy} where comparisons between different nuclear PDFs and study of theoretical uncertainties due to their uncertainties and scale variations are done. As a remark on experimental side, studies of direct photons in
the energy $\sqrt{s} = 8.16$~TeV at LHCb \cite{Boettcher:2019kxa} for $pPb$ and $Pbp$ are now well underway. They probe small $p_T$ region at very forward rapidities, which is ideal for investigating the nuclear effects in gluon sector.

\begin{figure*}[t]
\begin{tabular}{cc}
\includegraphics[scale=0.35]{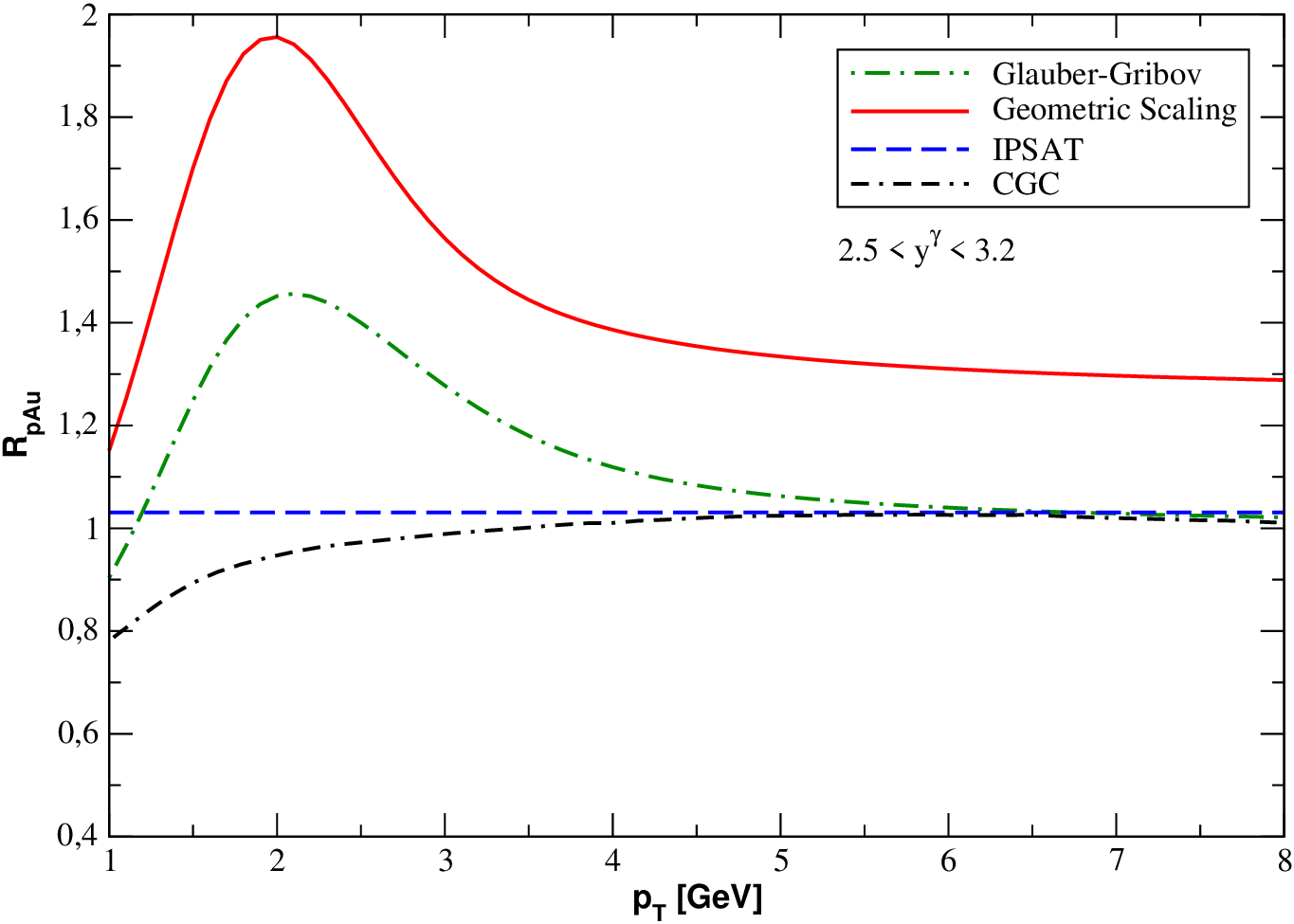}
\includegraphics[scale=0.35]{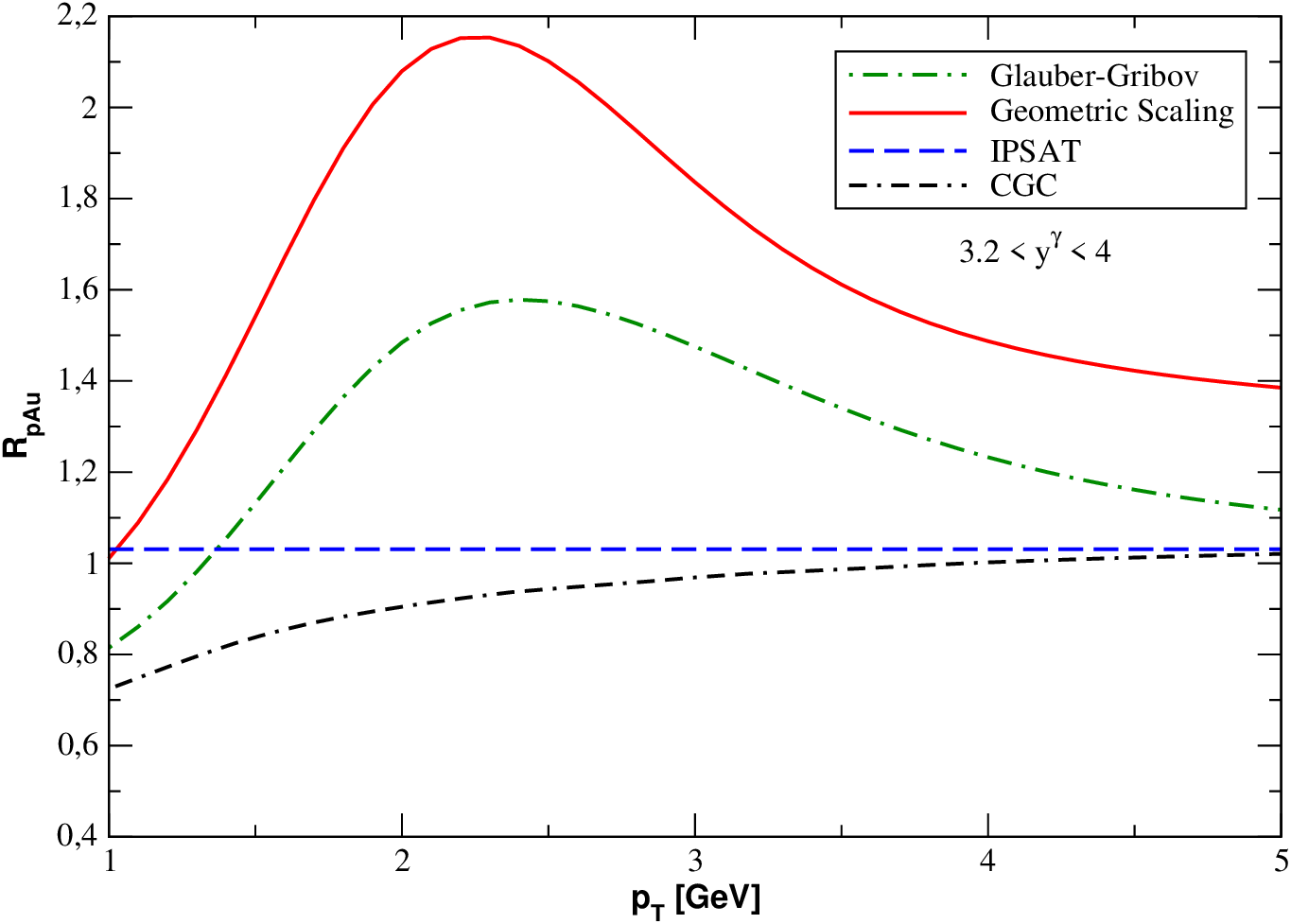}
\end{tabular}
\caption{Nuclear modification factor $R_{pAu}$ for prompt photon as a function of $p_T$ shown for two photon forward rapidity bins at
$\sqrt{s} = 200$~GeV. The predictions are obtained using GG, GS, and IPSAT approaches and compared to the results from CGC effective field theory.}
\label{compAu}
\end{figure*}

Additionally, for the purpose of continuing to make a comparison with predictions from CGC framework \cite{ducloue}, we show in Fig.~\ref{compAu} the results for the nuclear modification factor in $pAu$ collisions at RHIC at $\sqrt{s} = 200$~GeV and two photon forward rapidity bins: $2.5 < y^{\gamma} < 3.2$ and $3.2 < y^{\gamma} < 4$. Here, we found the same behavior pattern regarding the results as seen in $pPb$ case at $\sqrt{s} = 8$~TeV. Namely, $R_{pAu} < 2$ at small values of $p_T$ and points out that the nuclear effects are not perceptible towards larger $p_T$. However, the approaches predict less suppression at small $p_T$ in RHIC energy and reach the unity faster in comparison to the previous case. Interestingly, the GG and the CGC approach give similar results at sufficiently high $p_T$ in the two rapidity bins. The Cronin peak remains at low $p_T$ for GG and GS similarly to $pA$ collisions at the LHC.

\begin{figure*}[t]
\begin{center}
\includegraphics[scale=0.45]{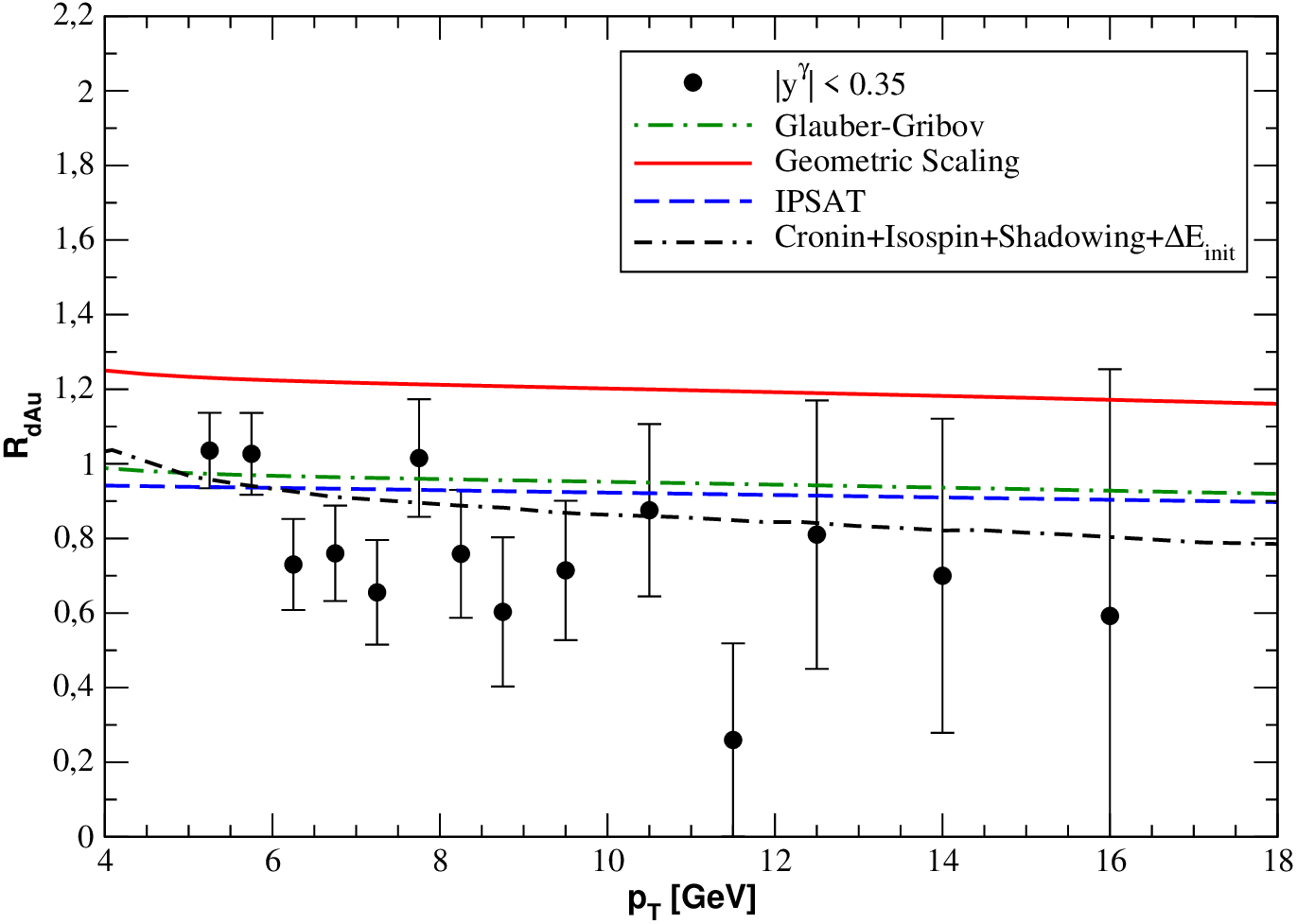}
\end{center}
\caption{Nuclear modification factor $R_{dAu}$ as a function of $p_T$ in midrapidity at $\sqrt{s}=200$~GeV. The predictions are obtained using GG, GS, and IPSAT approaches and compared to the pQCD calculations including CNM effects and to the experimental data from PHENIX Collaboration \cite{adare1}.}
\label{rdAu}
\end{figure*}

For sake of completeness, we present in Fig.~\ref{rdAu} the predictions considering the energy of $\sqrt{s}=200$~GeV at RHIC for minimum bias $dAu$ collisions at midrapidity. We include the experimental data for $R_{dAu}$ extracted by PHENIX Collaboration \cite{adare1}. The ratio is defined as
\begin{eqnarray}
R_{dAu}^{\gamma}= \frac{dN^{d+Au\rightarrow \gamma +X}/dyd^2\vec{p_T}}{\langle N_{\mathrm{coll}}\,\rangle \,dN^{p+p\rightarrow \gamma +X}/dyd^2\vec{p_T}},
\end{eqnarray}
where $\langle N_{\mathrm{coll}}\rangle$ is the average number of $NN$ binary collisions.

Predictions are compared to the calculations from Ref.~\cite{vitev} (dot-dashed line) with different combinations of initial-state effects (Cronin enhancement, isospin correction, nuclear effects embedded in nPDFs, and initial-state parton energy loss), i.e., cold nuclear matter effects (CNM). The GG and IPSAT results are not dependent on $p_T$ producing a constant ratio of order 0.9 and quite similar to the CNM results. The GS prediction presents the same pattern as in LHC energies and central rapidities, with $R_{pA}^{\gamma} \sim 1.3$. The main uncertainty in the GS approach is the prescription for the nuclear saturation scale, $Q_{s,A}^2$, and in Ref. \cite{gsds} we determined the uncertainty being $\sim 20\%$.

Finally, we present the results concerning the $x_T$-scaling observed in prompt photon production at central rapidities, $y^{\gamma}\approx 0$. Hence, the invariant cross section for inclusive particle production can be expressed as
\begin{eqnarray}
E \frac{d^{3}\sigma}{d^3p}=\frac{G(x_T)}{\left[\sqrt{s}\right]^{n_{\mathrm{eff}}(\sqrt{s},\,x_T)}}.
\end{eqnarray}
The scaling works for almost all the available data, with the power of the invariant cross section becoming softer towards higher $x_T$. The effective power is empirically determined as $n_{\mathrm{eff}}=4.5$ \cite{Vogelsang:1997cq,dEnterria:2012kvo,David:2019wpt}. The usual pQCD approach without hard scale evolution predicts the invariant cross section being proportional to $[\sqrt{s}]^4$ \cite{Vogelsang:1997cq}.

\begin{figure*}[t]
\begin{tabular}{cc}
\includegraphics[scale=0.35]{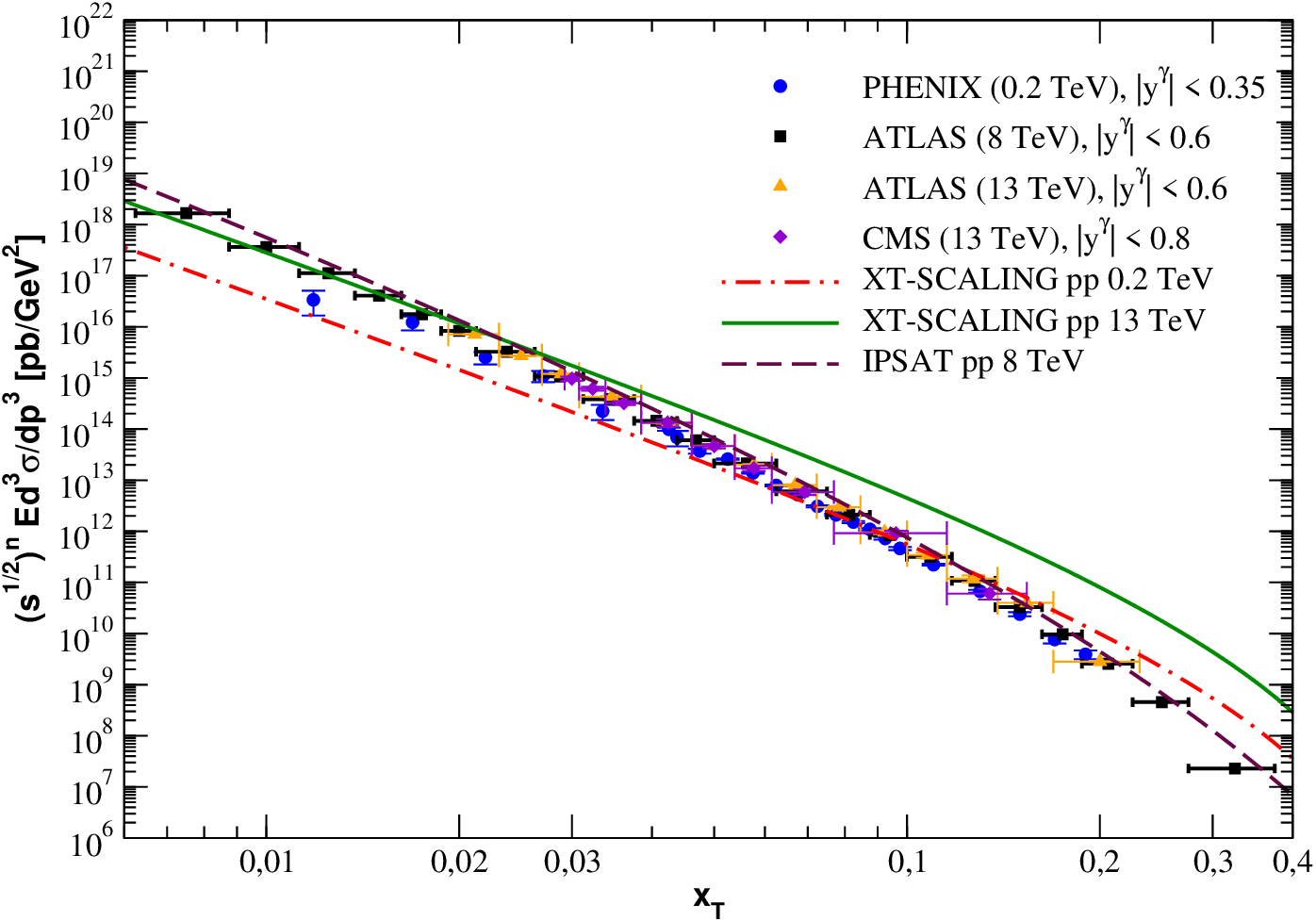}
\includegraphics[scale=0.35]{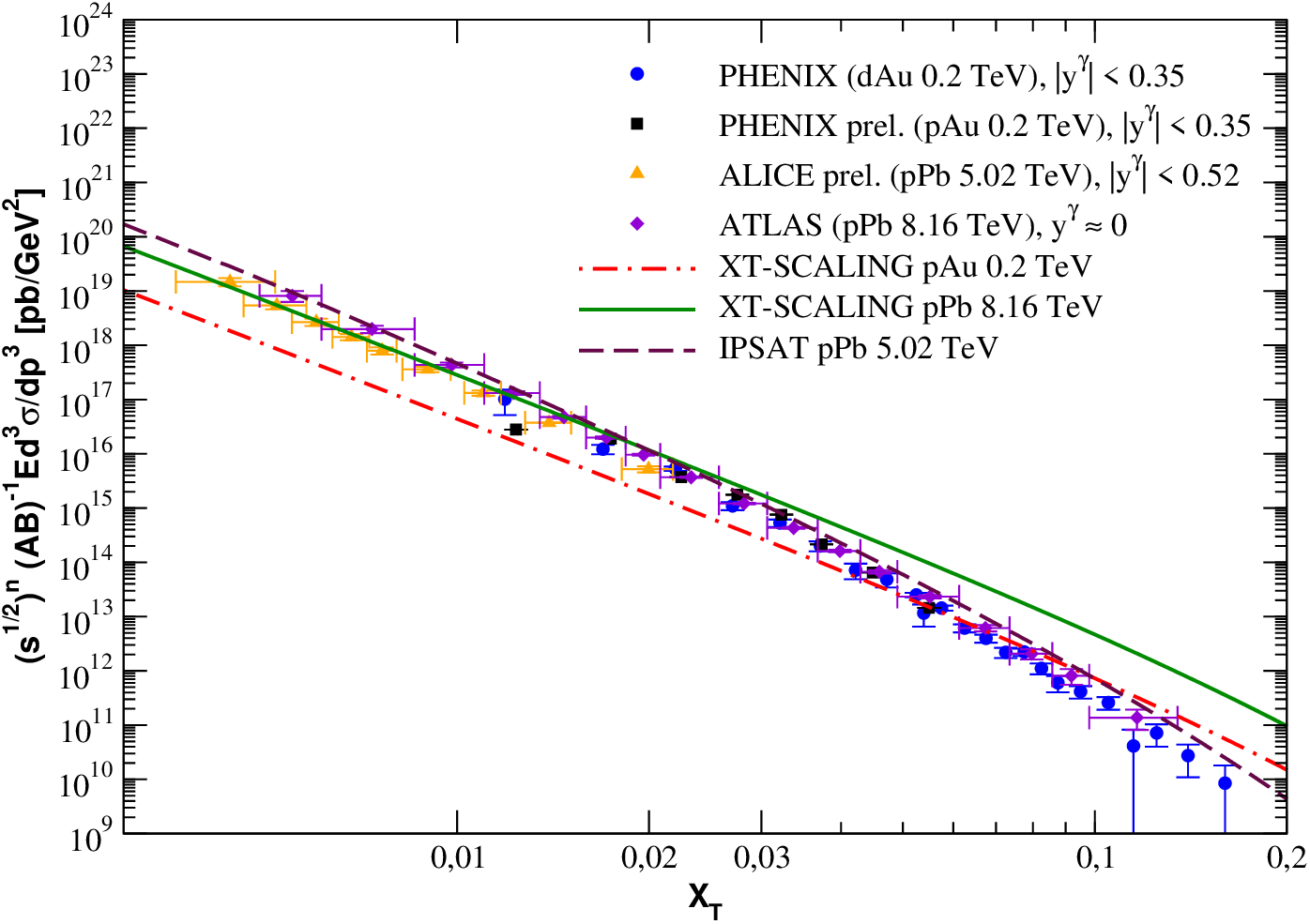}
\end{tabular}
\caption{The $x_T$-scaling of prompt photon production in $pp$ (left panel) and $dAu/pAu/pPb$ (right panel, normalized by nucleons number) collisions at midrapidity. Analytic expressions of Eqs.~(\ref{xtpp}) and (\ref{xtpA}) are presented in the limiting energies. The full calculation using the IPSAT model (including DGLAP evolution, color transparency approximation) is shown at fixed energies of $\sqrt{s}=8$~TeV ($pp$) and $\sqrt{s}=5.02$~TeV ($pA$).}
\label{xts}
\end{figure*}

For illustration, the scaled cross sections are presented for $pp$ \cite{adler,adler1} and $dAu$ \cite{adare1} collisions at RHIC and $pp/pPb$ collisions at LHC. Figure~\ref{xts} shows our predictions for the invariant cross sections in terms of $x_T = 2\,p_T/\sqrt{s}$ compared to the data collected for $y^{\gamma}\approx 0$ by PHENIX at $\sqrt{s}=200$~GeV, ATLAS at $\sqrt{s}=8$ and $13$~TeV \cite{aaboud,Aad:2016xcr} as well as the CMS data at $13$~TeV \cite{sirunyan}. In $pp$ case (left panel), the analytic scaling curve, Eq.~(\ref{xtpp}), is shown for the limiting energies of $\sqrt{s}=200$~GeV (dot-dashed line) and $\sqrt{s}=13$~TeV (solid line). The scaling curves have the correct shape at small $x_T$, however the correct transition to large $x_T$ is not achieved. As discussed before, this would be solved if the quantity $\gamma_s$ becomes $p_T$-dependent (as in the BUW dipole model). It is remarkable the good agreement with a full calculation using the IPSAT model and the experimental measurements for any $p_T$. It is presented for $\sqrt{s} =8 $ TeV (long-dashed line) and we verified low sensitivity to the energy value. 

The same procedure is followed in the proton(deuterium)-nucleus case. We have normalized the invariant cross sections by $AB$, i.e., we considered the spectra per nucleon. We explicitly present the results for the analytical expression in Eq.~(\ref{xtpA}) in the limiting energies of $\sqrt{s}=200$~GeV (dot-dashed line) and $\sqrt{s}=8.16$~TeV (solid line). The IPSAT result for $\sqrt{s}=5.02$~TeV is represented by the long-dashed curve. There is a clear resemblance between the two $pp$ and $p(d)A$ cases. The published data from ($dAu$) PHENIX \cite{adare1} and ($pPb$) ATLAS  \cite{aaboud1} are included. For sake of illustration, we also included the preliminary data from PHENIX for $pAu$ collisions ($\sqrt{s}=200$~GeV) \cite{CanoaRoman:2019fnj,Khachatryan:2018evz} as well as the preliminary ALICE $pPb$ data ($\sqrt{s}=5.02$~TeV) \cite{Blau:2019pna}.  

Interestingly, PHENIX Collaboration \cite{Khachatryan:2018evz} has recently investigated the scaling of the direct photon yield, integrated for $p_T\geq 1.0$~GeV, as a function of charged-particle multiplicity, $dN_{ch}/d\eta |_{\eta=0}$. It was demonstrated a direct photon excess yield at small $p_T$ in central $pAu$ collisions above $N_{coll}$ scaled baseline fit for proton-proton collisions. The Collaboration claims it may originate from an existing QGP droplets in small central systems, suggesting the presence of a transition point
between small and large systems. It would be interesting to address this question with the CD picture presented here.

\section{Summary} 
\label{conc}

In this work, we estimate the nuclear modification factor for prompt photon production at the RHIC and LHC energies considering distinct rapidity bins. We analyze the influence of nuclear effects in the transverse momentum distribution of prompt photons, correspondingly introduced by Glauber-Gribov, geometric scaling and IPSAT (color transparency approximation) models. Our results do not indicate a strong suppression due the saturation effects, and there are not any free parameter in the calculations. The experimental measurements of $R_{pA}$ in both RHIC and LHC colliders are consistent with unity within the experimental uncertainties at different values of rapidity. The models are in agreement to data, with some of them presenting Cronin enhancement at low $p_T$. We demonstrate that the CGC predictions are distinct of ours and this suggests that future experimental measurements on nuclear modification factor at forward rapidities may be performed to discriminate the models. 

Moreover, we have demonstrated that the parametrization proposed for the $x_T$-scaling of prompt photon production, considering proton and nuclear targets, works very well in describe the corresponding experimental measurements at low transverse momentum. This is notable given the simplicity of the parametrizations obtained within the QCD color dipole formalism in the massless quark limit, which can be useful in data analysis of future experimental measurements of prompt photons.

\section*{Acknowledgements}

This work was partially financed by the Brazilian funding agencies CAPES, CNPq, and FAPERGS. This study was financed in part by the Coordena\c{c}\~ao de Aperfei\c{c}oamento de Pessoal de N\'{\i}vel Superior - Brasil (CAPES) -
Finance Code 001.




\begin{thebibliography}{99}

\bibitem{EIC} A.~Accardi {\it et al.}, Eur.\ Phys.\ J.\ A {\bf 52}, 268 (2016).

\bibitem{arsene} I.~Arsene {\it et al.} [BRAHMS Collaboration], Phys. Rev. Lett. {\bf 93}, 242303 (2004).

\bibitem{adams} J.~Adams {\it et al.} [STAR Collaboration], Phys. Rev. Lett. {\bf 97}, 152302 (2006).

\bibitem{adare} A.~Adare {\it et al.} [PHENIX Collaboration], Phys. Rev. Lett. {\bf 107}, 172301 (2011).

\bibitem{Goncalves:2020tvh}
V.~P.~Goncalves, Y.~Lima, R.~Pasechnik and M.~Sumbera,
Phys. Rev. D \textbf{101}, no.9, 094019 (2020).

\bibitem{JalilianMarian:2012bd}
J.~Jalilian-Marian and A.~H.~Rezaeian,
Phys. Rev. D \textbf{86}, 034016 (2012).

\bibitem{amir} A.~H.~Rezaeian and A.~Schafer, Phys.\ Rev.\ D {\bf 81}, 114032 (2010).

\bibitem{pasechnik} R.~Pasechnik and M.~Sumbera, Universe {\bf 3}, 7 (2017).

\bibitem{acharya} S.~Acharya {\it et al.} [ALICE Collaboration], Phys.\ Rev.\ C {\bf 99}, 024912 (2019).

\bibitem{gordon} L.~E.~Gordon and W.~Vogelsang, Phys.\ Rev.\ D {\bf 50}, 1901 (1994).

\bibitem{frixione} S.~Frixione, Phys.\ Lett.\ B {\bf 429}, 369 (1998).

\bibitem{helenius1} I.~Helenius, K.~J.~Eskola and H.~Paukkunen, JHEP {\bf 09}, 138 (2014).
    
\bibitem{kop1} B.~Z.~Kopeliovich, A.~Schafer and A.~V.~Tarasov, Phys.\ Rev.\ C {\bf 59}, 1609 (1999).

\bibitem{kop2} B.~Z.~Kopeliovich, proc.\ of the workshop Hirschegg '95:
Dynamical Properties of Hadrons in Nuclear Matter, Hirschegg January
16-21, 1995, ed. by H. Feldmeyer and W. N\"orenberg, Darmstadt, 1995, p. 102 (hep-ph/9609385).

\bibitem{kop} B.~Z.~Kopeliovich, A.~H.~Rezaeian, H.~J.~Priner and I.~Schmidt, Phys. Lett. B {\bf 653}, 210 (2007).

\bibitem{munier} S.~Munier and R.~Peschanski, Phys.\ Rev.\ Lett.\ {\bf91}, 232001 (2003).

\bibitem{gsds} G.~Sampaio~dos~Santos, G.~Gil~da~Silveira and M.~V.~T.~Machado, arXiv:2004.02686 [hep-ph].

\bibitem{kop3} B.~Z.~Kopeliovich, J.~Raufeisen, A.~V.~Tarasov and M.~B.~Johnson, Phys.\ Rev.\ C {\bf 67}, 014903 (2003).

\bibitem{adeva} B.~Adeva {\it et al.}, Phys.\ Rev.\ D {\bf 58}, 112001 (1998).

\bibitem{kop4} A.~B.~Zamolodchikov, B.~Z.~Kopeliovich and L.~I.~Lapidus, JETP Lett. {\bf 33}, 595 (1981).

\bibitem{gbw} K.~J.~Golec-Biernat and M.~Wusthoff, Phys.\ Rev.\  D {\bf 59}, 014017 (1998).

\bibitem{gbwfit} K.~Golec-Biernat and S.~Sapeta, JHEP {\bf 1803}, 102 (2018).

\bibitem{ipsat} H.~Kowalski and D.~Teaney, Phys.\ Rev.\ D {\bf 68}, 114005 (2003).

\bibitem{ipsatfit} A.~H.~Rezaeian, M.~Siddikov, M.~Van de Klundert and R.~Venugopalan, Phys.\ Rev.\ D {\bf 87}, 034002 (2013).  

\bibitem{mm} M.~V.~T.~Machado and C.~B.~Mariotto, Eur.\ Phys.\ J.\ C {\bf 61}, 871 (2009).

\bibitem{salgado} N.~Armesto, C.~A.~Salgado and U.~A.~Wiedemann, Phys.\ Rev.\ Lett.\  {\bf 94}, 022002 (2005).

\bibitem{ben} F.~G.~Ben, M.~V.~T.~Machado and W.~K.~Sauter, Phys.\ Rev.\ D {\bf 96}, 054015 (2017).

\bibitem{armesto} N.~Armesto, Eur. Phys. J. C {\bf 26}, 35 (2002).

\bibitem{buw} D.~Boer, A.~Utermann and E.~Wessels, Phys.\ Rev.\ D {\bf 77}, 054014 (2008).

\bibitem{kbml} C.~Klein-Bösing and L.~McLerran, Phys.\ Lett.\ B {\bf 734}, 282 (2014).
  
\bibitem{pras} M.~Praszałowicz, EPJ Web Conf.\ {\bf 206}, 02002 (2019).
  
\bibitem{kp} V.~Khachatryan and M.~Praszalowicz, arXiv:1907.03815 [nucl-th].

\bibitem{allm} D.~Gabbert and L.~De~Nardo, arXiv:0708.3196 [hep-ph].

\bibitem{aaboud1} M.~Aaboud {\it et al.} [ATLAS Collaboration], Phys.\ Lett.\ B {\bf 796}, 230 (2019).

\bibitem{jetphox} P.~Aurenche, J.~P.~Guillet, E.~Pilon, M.~Werlen, M.~Fontannaz, Phys.\ Rev.\ D {\bf 73} 094007 (2006).

\bibitem{kovarik} K.~Kovarik {\it et al.}, Phys.\ Rev.\ D {\bf 93}, 085037 (2016).

\bibitem{ducloue} B.~Duclou\'e, T.~Lappi and H.~M\"antysaari, Phys.\ Rev.\ D {\bf 97}, 054023 (2018).

\bibitem{Arleo:2007js}
F.~Arleo and T.~Gousset,
Phys. Lett. B \textbf{660}, 181-187 (2008).

\bibitem{Goharipour:2017uic}
M.~Goharipour and H.~Mehraban,
Phys. Rev. D \textbf{95}, no.5, 054002 (2017).

\bibitem{Goharipour:2018sip}
M.~Goharipour and S.~Rostami,
Phys. Rev. C \textbf{99}, no.5, 055206 (2019).

\bibitem{Klasen:2017dsy}
M.~Klasen, C.~Klein-Bösing and H.~Poppenborg,
JHEP \textbf{03}, 081 (2018).


\bibitem{Boettcher:2019kxa}
T.~Boettcher [LHCb],
Nucl. Phys. A \textbf{982}, 251-254 (2019).


\bibitem{adare1} A.~Adare {\it et al.} [PHENIX Collaboration],  Phys.\ Rev.\ C {\bf 87}, 054907 (2013).

\bibitem{vitev} I.~Vitev and B.~W.~Zhang, Phys.\ Lett.\ B {\bf 669}, 337 (2008).

\bibitem{Vogelsang:1997cq}
W.~Vogelsang and M.~Whalley,
J. Phys. G \textbf{23}, A1-A69 (1997).

\bibitem{dEnterria:2012kvo}
D.~d'Enterria and J.~Rojo,
Nucl. Phys. B \textbf{860}, 311-338 (2012).

\bibitem{David:2019wpt}
G.~David,
Rept. Prog. Phys. \textbf{83}, no.4, 046301 (2020).

\bibitem{adler} S.~S.~Adler {\it et al.} [PHENIX Collaboration], Phys.\ Rev.\ Lett.\ {\bf 98}, 012002 (2007).

\bibitem{adler1} S.~S.~Adler {\it et al.} [PHENIX Collaboration], Phys.\ Rev.\ D {\bf 86}, 072008 (2012).

\bibitem{aaboud} M.~Aaboud {\it et al.} [ATLAS Collaboration], Phys.\ Lett.\ B {\bf 770}, 473 (2017).

\bibitem{Aad:2016xcr}
G.~Aad \textit{et al.} [ATLAS],
JHEP \textbf{08}, 005 (2016).

\bibitem{sirunyan} A.~M.~Sirunyan {\it et al.} [CMS Collaboration], Eur.\ Phys.\ J.\ C {\bf 79}, 20 (2019).

\bibitem{CanoaRoman:2019fnj}
V.~Canoa Roman,
MDPI Proc. \textbf{10}, no.1, 32 (2019).

\bibitem{Khachatryan:2018evz}
V.~Khachatryan [PHENIX],
Nucl. Phys. A \textbf{982}, 763-766 (2019).

\bibitem{Blau:2019pna}
D.~Blau [ALICE],
EPJ Web Conf. \textbf{222}, 02001 (2019).

\end{thebibliography}
\end{document}